\documentstyle[twocolumn,prl,aps,epsf]{revtex}

\def\beqn{\begin{eqnarray}}
\def\eeqn{\end{eqnarray}}
\def\beqns{\begin{eqnarray*}}
\def\eeqns{\end{eqnarray*}}
\def\beq{\begin{equation}}
\def\eeq{\end{equation}}
\def\bea{\begin{array}}
\def\ea{\end{array}}

\def\<{\langle}
\def\>{\rangle}
\def\ham{\hat{H}}

\draft
\widetext

\begin{document}

\twocolumn[\hsize\textwidth\columnwidth\hsize\csname
 @twocolumnfalse\endcsname
\draft
\preprint{}

\title{Anomalous Quantum Diffusion and Conductivity of Quasicrystals}

\author{Didier Mayou}

\address{ LEPES-CNRS, 25 avenue des Martyrs BP166, 38042 
Grenoble, France.}

\maketitle

\begin{abstract}
\leftskip 54.8pt
\rightskip 54.8pt

A phenomenological diffusion law $L(t)\propto t^{\beta}$, where
$L(t)$ measures the spreading of a wave-packet in a time $t$, is
assumed for perfect quasicrystals. We show that it affects their
conductivity  with striking differences compared to the
case of periodic metals. In the absence of defects  the dissipative
part of conductivity is non zero  even at low frequencies contrary to
the case of crystals. Also if $\beta<1/2$ the d.c. conductivity increases when
disorder increases and the so-called Drude peak, characteristic of metals, is
replaced  by a dip . Experimental results are briefly discussed.

\end{abstract}


]

\noindent

Since the pioneering work of Kohmoto et al \cite{Kohmoto1}
numerous studies of quasiperiodic Hamiltonians have given evidences that the
associated eigenstates are multifractal and critical with an  algebraic decrease of
their envelop at large distances \cite{Kohmoto2,Schreiber,Fujiwara1}. These states
lead to a peculiar quantum diffusion which is neither ballistic as in crystals nor
diffusive as in metallic disordered systems but follows at sufficiently large time a
power law  $L(t)\propto t^{\beta}$ where $L(t)$ measures the spreading of a
wave-packet in a time $t$ \cite {Piechon,Sire1,Mantica}. $\beta$ depends on the energy
of the wave-packet and on the Hamiltonian but is restricted to $0 \leq \beta \leq 1$. For
realistic models, periodic approximants of quasicrystals
\cite{Fujiwara2,Fujiwara3,Hafner} have narrow bands, in agreement with the tendency to
localization. For some phases these computations have even confirmed a multifractal
character of eigenstates\cite{Fujiwara2}. Electronic transport has been studied through
Bloch-Boltzmann type approximation assuming a peculiar bandstructure
\cite{Fujiwara3,Hafner,Ashcroft} and this provides interesting information. Yet this
approach does not consider the anomalous diffusion.

\noindent This paper presents a formalism that allows to discuss, for the first time to
our knowledge, how the a.c. conductivity is affected by the anomalous
diffusion. For d.c. conductivity of quasiperiodic systems with some disorder, the role
of anomalous diffusion has been considered by several authors 
\cite{Mayou,Sire2,Bellissard}. Using the Einstein formula  $\sigma = 2
e^{2}n(E_{F})D(E_{F})$ where $n(E_{F})$ is the density of states at the Fermi energy
per spin and $D(E_{F})$ the diffusivity of states at the Fermi energy one expects that
the zero frequency conductivity is :

\beq
\sigma_{DC}(\tau) \simeq 2 e^{2}n(E)A\tau^{2\beta-1}
\label{A}
\eeq

 where $\tau $ is the time beyond which the propagation becomes
diffusive due to disorder. In this scheme the propagation on time scale $t>\tau$  is
diffusive with $D(E_{F}) \simeq
\frac{L^{2}(\tau)}{3\tau}\simeq  A\tau ^{2\beta-1}$. For $\beta<1/2$ the
conductivity increases with disorder as observed in some quasicrystalline alloys 
\cite{Berger}. 
From a dimensional argument one expects that the frequency dependent conductivity
$\sigma_{AC}(\omega) = B\sigma_{DC}(\tau = \frac{1}{\omega})\propto
\frac{1}{\omega^{2\beta-1}}$ at the lowest frequencies where $B$ is some complex
number with a modulus of order unity . The real part of $B$ is related to energy
dissipation by the system. For crystals it is zero but the dimensional argument alone
cannot give its value in quasicrystals.

\noindent In this paper we study the zero or low frequency conductivity of
a quasiperiodic Hamiltonian of independent electrons with and without static disorder.
It is shown first that a suitably defined average
of the conductivity at frequency $\omega$ is related to a quantity
$\tilde{\sigma}(\omega)$ which has the analytical properties of a conductivity and is
directly related to quantum diffusion. The behaviour of $\tilde{\sigma}(\omega)$ with
frequency or disorder is thus representative of that of the conductivity and is studied in
detail. Assuming an anomalous diffusion law in the perfect quasicrystal we obtain an
exact result for the conductivity $\tilde{\sigma}_{Pure}(\omega)$ of the quasiperiodic system
without defects. Then we consider the role of static disorder and introduce a relaxation
time approximation for the velocity auto-correlation function which physical meaning
and limits are discussed. This approximation allows to derive a generalized Drude formula
for  $\tilde{\sigma}(\omega)$ that interpolates between the above $\sigma_{DC}(\tau)$
(\ref{A}) and $\tilde{\sigma}_{Pure}(\omega)$ as a function of the parameter $\omega\tau$. This
formula implies a drastic deviation of optical conductivity from the case of crystals. We
comment shortly on the conductivity of several quasicrystalline phases.

\noindent Let us define several quantities  and give some exact relations
between these quantities. We define first $<f>_{E}$, the average diagonal element
of an operator $f$ among states at energy  $E$  : 

\beq
<f>_{E}= \frac{Tr[\delta(E-\ham) f]}{Tr[\delta(E-\ham)]}
\label{Def 1}
\eeq

$Tr$ means a trace of an operator and $\delta(E-\ham)$ projects on eigenstates of the
Hamiltonian $\ham$ at energy $E$. Next we define a measure $X^{2}(E,t)$ of the
spreading of wave-packets at energy $E$ along the x-direction :  

\beq
X^{2}(E,t) = <(X(t)-X(0))^{2}>_{E}
\label{Def 2}
\eeq

where $X(t)$ is the position operator along the x-direction  at time $t$  in the
Heisenberg representation. $X^{2}(E,t)$ is a real and pair function of time.
\noindent We define also the auto-correlation
function $C(E,t-t')$  for the velocity $V_{x}(t)= \frac{dX(t)}{dt}$ 
between  $t$ and $t'$ and which from (\ref{Def 2}) satisfies:

\beq
C(E,t-t') = <V_{x}(t)V_{x}(t') + V_{x}(t')V_{x}(t)>_{E}
\label{Def 3}
\eeq
\beq
\frac{dX^{2}(E,t)}{dt}=\int\limits_{0}^{t}dt'C(E,t')
\label{Def 4}
\eeq

$C(E,t)$ is a real and pair function of $t$. For pure quasicrystals the phenomenological
diffusion law reads  :

\beq
X_{0}^{2}(E,t)\simeq At^{2\beta}
\label{Def 5}
\eeq
\beq
C_{0}(E,t)= \frac{d^{2}X_{0}^{2}(E,t)}{dt^{2}}\simeq A2\beta(2\beta -1)t^{2\beta -2}
\label{Def 6}
\eeq

(\ref{Def 6}) is obtained from (\ref{Def 4}) and (\ref{Def 5}). In periodic approximants,
the scaling law $X_{0}^{2}(E,t)\simeq At^{2\beta}$ will be valid only up to some maximum time
$t_{2}$ beyond which the periodicty is effective. Then quantities deduced from the
scaling law (see below) will be valid for time scales $\tau\leq t_{2}$ or
frequencies scales $\omega \geq \frac{1}{t_{2}}$.  The Guarneri inequality \cite{Guarneri}
implies a singular continuous density of states of a three dimensional quasicrystal for
$\beta <1/3 $. Yet if we consider periodic approximants we can assume that the density of
states is absolutely continuous even for $\beta < 1/3 $ but with the above restriction on
time and frequency scales. From its definition (\ref{Def 3}) $C(E,t=0) = 2<V_{x}^{2}>_{E}$
is finite which implies that the propagation is ballistic at the lowest times. Thus the
anomalous diffusion law can be valid only at sufficiently large times $t>t_{1}$. It
occurs only when the wave-packet has sufficient time to ''feel'' specific atomic
arrangements \cite{Trambly,Janot} and longer range quasiperiodic order. This remark is
important for the sum-rule satisfied by conductivity (\ref{Relaxation 4}). 

\noindent We
introduce $F(E,\omega)$ that appears in the formula \cite{Kubo} for the
component $\sigma_{x,x}(\mu,\omega)$ of the conductivity tensor and
satisfies: 

\beq F(E,\omega) =\frac{2\pi\hbar
e^{2}}{\Omega} Tr[\delta(E-\ham)V_{x}\delta(E+\hbar \omega -\ham)V_{x}]
\label{Def 7}
\eeq
\beq
S(E,\omega) = F(E,\omega) + F(E -\hbar\omega,\omega)
\label{Def 8A}
\eeq
\beq
S(E,\omega) =
e^{2}n(E)\int\limits_{-\infty}^{+\infty}exp(i\omega t) C(E,t)dt
\label{Def 8B}
\eeq
$\Omega$ is the volume of the system. Starting from (\ref{Def 7}),  (\ref{Def 8B}) is
easily shown by using the Fourier expansion of $\delta(E-\ham)$ and the cyclic property
of a trace \cite{Mayou2}. If $F(E,\omega)$  depends
wealky on the energy $E$ at a given $\omega$ then 
$F(E,\omega)\simeq F(E-\hbar\omega,\omega) \simeq S(E,\omega)/2$
which gives $F(E,\omega)$ in terms of the velocity correlation
function. This relation is often valid in disordered systems at low frequency \cite{TVR}.
Yet in quasicrystals there are important variations of electronic structure on small
energy scale (tenths or hundredth of electron Volt and possibly less \cite
{Fujiwara2,Fujiwara3}) and the above assumption could have a vanishingly small range
of validity  . Since we are interested in general trends we consider an
average of the conductivity at frequency $\omega$ over a range of chemical potential
values (we could also consider the conductivity at very high temperature
$kT>>\hbar\omega$). More precisely we define averages of functions $g(E)$ by a
convolution with a positive function $P(x)$  such that $\int dx P(x) =1$ :

\beq
<g(E)>_{P} = \int dx P(x) g(E-x)
\label{KG0}
\eeq

 One can choose for example $ P(x) = \frac{1}{2W}
exp(-\frac{|x|}{W})$ where $W$ is its width. From the  Kubo-Greenwood formula \cite{Kubo}
at $T=0$ :

\beq
<Re\sigma_{x,x}(\mu,\omega)>_{P} = \int\limits_{\mu -\hbar\omega}^{\mu}
\frac{dE}{\hbar\omega}  <F(E,\omega)>_{P} \label{KG1}
\eeq
where $Re$ means the real part. If $W>> \hbar\omega$
  $F(E,\omega)$ and $F(E -\hbar\omega,\omega)$
have nearly the same weight in $<F(E,\omega)>_{P}$ (\ref{KG0}). Then only their sum
$S(E,\omega)$ is relevant and $F(E,\omega)$ can be replaced by $\frac{S(E,\omega)}{2}$ in
(\ref{KG1}). One can write :

\beq
<Re\sigma_{x,x}(\mu,\omega)>_{P}\simeq \int \limits_{\mu -\hbar\omega}^{\mu}
\frac{dE}{\hbar\omega}
<Re\tilde{\sigma}(E,\omega)>_{P} 
\label{KG3}
\eeq
\beq
\tilde{\sigma}(\mu,\omega) = e^{2}n(\mu)
\int\limits_{0}^{\infty}exp(i\omega t) C(\mu,t)dt
\label{Sigma t}
\eeq

since for $\omega$ real  $S(E,\omega) = 2 Re \tilde{\sigma}(E,\omega)$. The relative
difference between the two members of (\ref{KG3}) is bounded by a term of order
$\frac{\hbar\omega}{W}$ for $\omega$ real \cite{Mayou2}. $\tilde{\sigma}(\mu,\omega)$ has
the dimension of a conductivity. It is an analytical function of $\omega$ in the upper
half of the complex plane. For large $\omega$  $\tilde{\sigma}(\mu,\omega) \propto
\frac{1}{\omega}$ and the Kramers-Kr\"onig relations can be proved. Finally the usual sum
rule is also valid : 

\beq
\int\limits_{0}^{\infty}Re\tilde{\sigma}(\mu,\omega)d\omega=\pi
e^{2}n(\mu) \frac{C(\mu,t=0)}{2} = \pi e^{2}\frac{n}{2m^{*}}
\label{Relaxation 4}
\eeq
where the last equality holds for free electron like systems with electron density $n$
and effective mass $m^{*}$. From now on we focus on the properties of
$\tilde{\sigma}(\mu,\omega)$ since  (\ref{KG3}) shows that its variation with frequency
or with  defects drives the variation of the average conductivity
$<\sigma(\mu,\omega)>_{P}$. Due to Kramers-Kr\"onig relations we expect also that the
imaginary part of $\tilde{\sigma}(\mu,\omega)$ is representative of that of the averaged
conductivity $<\sigma(\mu,\omega)>_{P}$ as discussed in \cite{Mayou2}.

\noindent Using the  velocity correlation (\ref{Def 6}) associated to the anomalous
 diffusion law (\ref{Def 5}) we deduce  
$\tilde{\sigma}_{Pure}(\mu,\omega)$  :

\beq
\tilde{\sigma}_{Pure}(\mu,\omega) \simeq e^{2}n(\mu)A \Gamma(2\beta +1)
(\frac{i}{\omega})^{2\beta-1}
\label{Drude 1}
\eeq

since the anomalous diffusion is valid only at $t>t_{1}$  the above formula is
valid for $\omega \leq\frac{1}{t_{1}}$.  $\Gamma(z)$ is the Gamma function
and is of order one for $0\leq \beta \leq 1$ \cite{handbook}. 

 \noindent The real part of this expression is strictly positive if $0<\beta<1$ since
 $Re(i^{2\beta-1})=cos(\frac{\pi}{2}(2\beta-1))$ . Thus there is dissipation even for
a perfect quasiperiodic system. This is in contrast with the case of crystals
$(\beta =1$) for which the dissipation is zero.  A natural interpretation
is obtained by comparing the bandstructure of crystals and that of approximants of
quasicrystals in the limit of infinite unit cell. In a crystal the number of bands is
finite and when $\omega$ tends to zero there is no more possibility
for vertical interband transition at energy $\hbar\omega$ between an occupied state
and an empty one. Since the system cannot absorb energy the real part of the
conductivity is zero. On the contrary the number of bands of approximants with large
unit cells increases indefinitely with the size of the unit cell. In the limit of
infinitely large  unit cell interband transition are possible even at the smallest
frequencies and the dissipative part of conductivity is non zero.

\noindent We consider now the effect of static disorder. For weak
static disorder numerical studies \cite {Roche} give evidences that the propagation is
diffusive on time scale $t>\tau$. There could be exceptions \cite{Triozon}
at very strong quasiperiodic potential for which the quantum diffusion stays
anomalous even with static disorder. In this case the result (\ref{Drude 1}) is
still valid. Our aim is to find an interpolation formula from the low frequency regime 
$\omega\tau <<1$ (\ref{A}) to the high frequency regime $\omega\tau>>1$ (\ref{Drude 1}) .
Let us  assume a  relaxation time  approximation for $C(E,t)$  :

\beq
C(E,t) \simeq C_{0}(E,t) exp(-\frac{|t|}{\tau})
\label{Relaxation 1}
\eeq

where $C_{0}(E,t)$ is the velocity correlation function of the system without
disorder. The physical meaning is that due to disorder the electron is scattered in
states with no or smaller velocity correlation leading to a damping of the
velocity correlation function. Using (\ref{Relaxation 1}) we see that in (\ref{Def 4})
$\frac{dX^{2}(E,t)}{dt}$ is constant, and the propagation is
diffusive, for $t>\tau$. The approximation (\ref{Relaxation 1}) could be generalized by replacing the
exponential by an average of damped exponentials. It will be clear that due to the
linearity of equations this does not change our qualitative conclusions.

\noindent Assuming (\ref{Relaxation 1}), using  $X_{0}^{2}(E,0) = 0 $ 
(see(\ref{Def 2})), $\frac{dX_{0}^{2}(E,t)}{dt}= 0$ at $t=0$ (see (\ref{Def 4})) and 
defining $z=(\frac{1}{\tau} - i\omega)$, $\tilde{\sigma}(\mu,\omega)$ (\ref{Sigma t}) is
equivalent to :

\beq
\tilde{\sigma}(\mu,\omega) \simeq e^{2}n(\mu)A z ^{2}
\int\limits_{0}^{\infty}dt X_{0}^{2}(E,t) exp(-zt)
\label{Drude 2}
\eeq
\beq
\tilde{\sigma}(\mu,\omega)  \simeq e^{2}n(\mu)A \Gamma(2\beta +1)
(\frac{\tau}{1-i\omega\tau})^{2\beta-1}
\label{Drude 3}
\eeq

\noindent (\ref{Drude 2}) expresses a direct relation between
quantum diffusion and optical conductivity. (\ref{Drude 3}) is obtained from 
(\ref{Drude 2}) and (\ref{Def 5}). Since the
diffusion law (\ref{Def 5}) is valid  for $t>t_{1}$ (\ref{Drude 3}) applies only for
$\omega<\frac{1}{t_{1}}$ and $\tau>t_{1}$  and this generalized Drude formula does not
satisfiy the sum rule (\ref{Relaxation 4}). The point is that it interpolates between
(\ref{Drude 1}) for $\omega\tau >> 1$ and (\ref{A}) for  $\omega\tau <<1$ (up to a
numerical factor of order one).

\begin{figure}[htbp]
\epsfxsize=8cm
\centerline{\epsffile{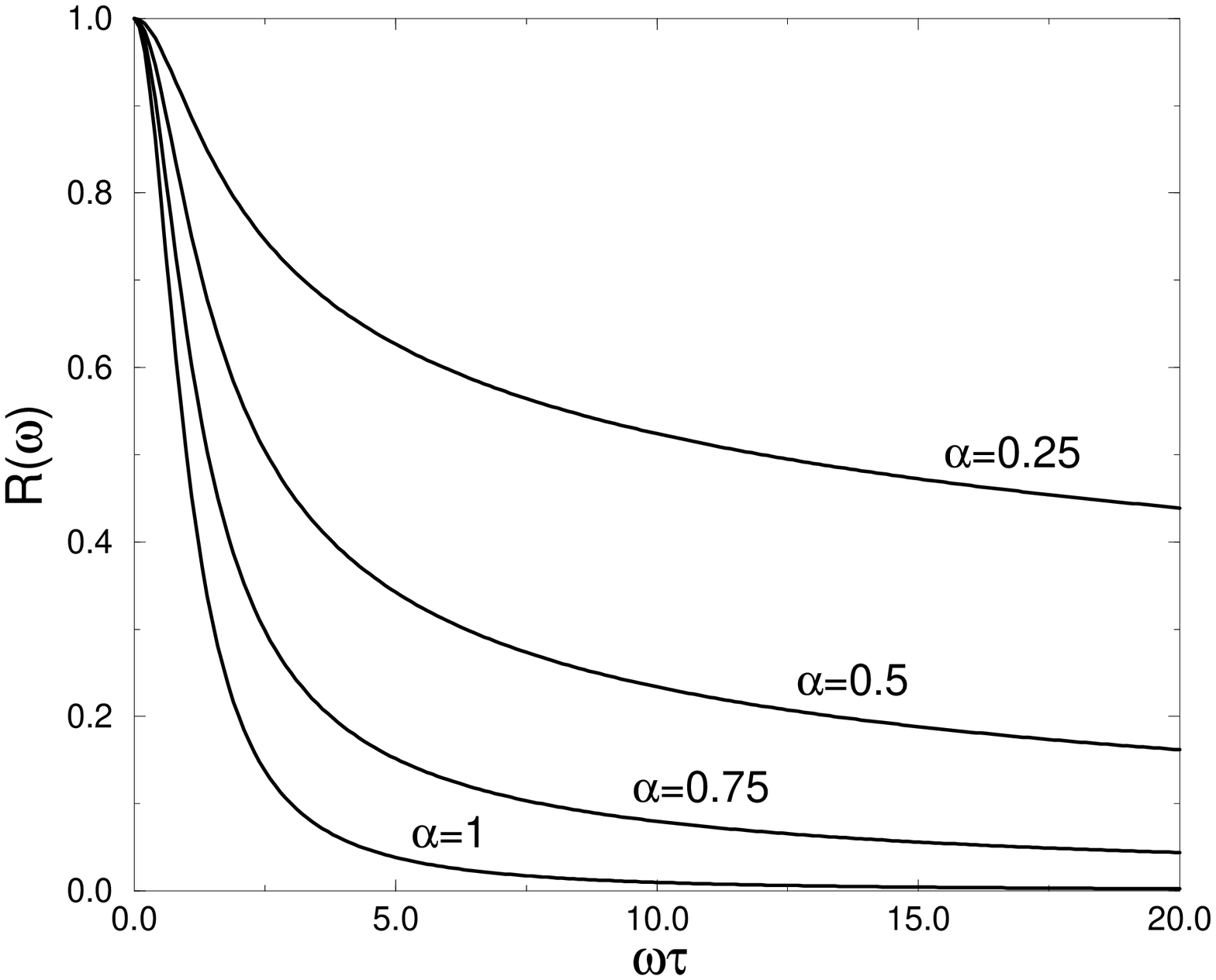}}
\epsfxsize=8cm
\centerline{\epsffile{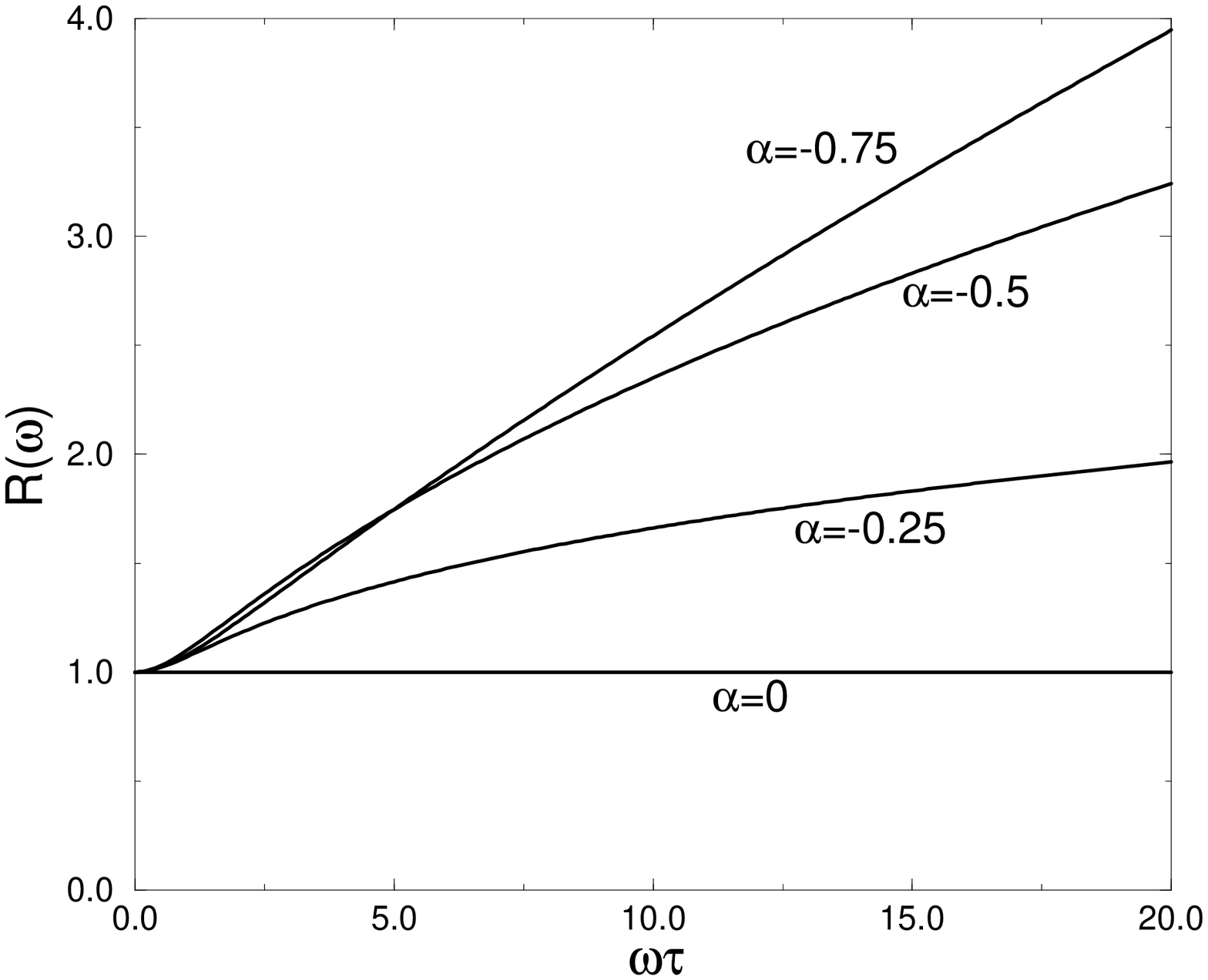}}
\caption{Variation of
$R(\omega)=$ $\displaystyle{Re\tilde{\sigma}(\mu,\omega)\over
\tilde{\sigma}(\mu,0)}$ as a function of $\omega\tau$ for several values of
$\alpha=(2\beta -1)$} \label{fig beta}
 \end{figure}

Figure (\ref{fig beta}) shows 
$R(\omega)=$ $\displaystyle{Re\tilde{\sigma}(\mu,\omega)\over \tilde{\sigma}(\mu,0)}$
as a function of $\omega \tau$.  For $\alpha =2 \beta -1 >0 $ the so-called Drude
peak is less marked than in crystals yet the distinction may be difficult to
establish experimentally. The deviation from the case of crystals is drastic if
$\alpha =2 \beta -1 \leq 0 $   : instead of a Drude peak there is a plateau or a dip. 

\noindent It is important to note that  due to quantum
interferences similar to those of disordered systems \cite{Mayou} the velocity
correlation function should become negative on time scale $t>\tau$. This effect is
beyond the relaxation time approximation (\ref{Relaxation 1}) and  becomes quantitatively important close
to the metal-insulator transition \cite{TVR}. According to the scaling theory of
localization the parameter $g = \sigma_{Drude} X_{0}(\tau)$ must be compared to the
universal critical conductance for three dimensional systems $g_{c}$. If $g>> g_{c}$
the corrections are negligible and the relaxation time approximation is essentially
correct. If $g$ is comparable to  $g_{c}$ the relaxation time approximation is
incorrect and if $g< g_{c}$ the quantum interferences even lead to an insulating
state. Quantum interferences are effective on time scale $t>\tau$ or $\omega
\leq \frac{1}{\tau}$ and it has been argued \cite{Mayou} that the system behave like a
standard disordered system at these time scales. In any case one expects that  
(\ref{Drude 3}) is correct for $\frac{1}{t_{1}}\geq \omega \geq \frac{1}{\tau}$.
Even though the formalism is for static disorder, we believe that the generalized Drude
formula (\ref{Drude 3}) is better for an inelastic scattering since 
then quantum interferences cannot appear.

\noindent  For many decagonal phases d-AlCuCo and icosahedral phases i-AlCuFe it
seems that $\frac{\hbar}{\tau(T)} \simeq  3 10^{-2}eV$ for $T\leq 400K$
\cite{Berger} then the full range in Figure (\ref{fig beta}) is
$\hbar\omega\simeq 0.6 eV $.  A detailed comparison of the present theory with experiment
is difficult since we consider only averages of the conductivity and also because
quantum interferences effects \cite{Berger} that are not included in the relaxation time
approximation should play a role at low frequency ($\omega \tau(T) \leq 1$). Yet since 
i-AlCuFe \cite{i-AlCuFe} and i-AlPdRe \cite{i-AlPdRe} do not exhibit a Drude peak they
are similar to the case  $2\beta-1$ negative. Correlatively (in agreement with (\ref{Drude
3})) the d.c. conductivity increases with temperature and probably also with the addition
of defects. For d-AlCuCo \cite{d-AlCuCo}, which consists of quasiperiodic planes stacked
periodically in the z-direction, the dissipative part of conductivity along the
quasiperiodic planes depends weakly on frequency which is consistent with $2\beta-1$
close to zero. Correlatively the conductivity in the quasiperiodic planes depends weakly
on temperature. Along the z direction the conductivity of d-AlCuCo is that of good metals.

\noindent To conclude let us recall that a suitably defined average of the conductivity
can be related to the quantum diffusion in the system $X_{0}(\mu,t)$ as expressed by
(\ref{KG3}, \ref{Drude 2}). Striking differences with crystals occur when $X_{0}(\mu,t)$
deviates from a ballistic law and this seems to be in agreement with experimental
results.

\noindent I would like to thank B.K. Chakraverty, F.Triozon, T.
Grenet, C. Berger and P. Lefebvre for discussions.

\vfill\eject

\end{document}